\documentclass[twocol]{arXiv}
\usepackage{sidecap}
\sidecaptionvpos{figure}{t}
\renewcommand{\thefigure}{{\bf \arabic{figure}}}
\renewcommand{\thetable}{{\bf \arabic{table}}}

\usepackage[
    papersize={8.5in,11in},
    twoside=false,
    inner=1.7cm, outer=1.7cm,
    top=2.5cm, bottom=1cm,
    footskip=1cm,
    nomarginpar,
]{geometry}

\usepackage{graphicx}%
\usepackage{amsmath,amssymb,amsfonts}%
\usepackage[separate-uncertainty = true, multi-part-units=single]{siunitx}%

\title{Recent Weakening of the Global Radiative Feedback}

\authors{
    Senne Van Loon${}^{1,\ast}$\correspondingauthor{
        Senne Van Loon; senne.van\_loon@colostate.edu;
    },  
    Maria Rugenstein${}^{1}$, 
    Mark D. Zelinka${}^{2}$, and
    Timothy Andrews${}^{3,4}$
}

\affiliation{
    ${}^{1}$Department of Atmospheric Science, Colorado State University, Fort Collins, Colorado, USA\\
    ${}^{2}$Lawrence Livermore National Laboratory, Livermore, California\\
    ${}^{3}$Met Office Hadley Centre, Exeter, UK\\
    ${}^{4}$School of Earth and Environment, University of Leeds, Leeds, UK
}

\abstract{%
Earth's climate stability, characterized by the global radiative feedback parameter ($\lambda$), varies decadally due to changing surface temperature patterns. Recent variations in $\lambda$ are poorly understood as coordinated model simulations typically end in 2014. We apply a convolutional neural network trained on climate model simulations to observation-based surface temperature reconstructions to estimate variations in $\lambda$ up to 2025. We find that $\lambda$ reached a minimum (maximum stability) around the mid 1990s ($\lambda\simeq{\rm\bf -3\, Wm^{-2}/K}$), but has since weakened significantly ($\lambda\simeq{\rm\bf -2\, Wm^{-2}/K}$). We confirm these results with climate model simulations extended to 2022. The recent $\lambda$ weakening is not significantly affected by El Ni\~{n}o Southern Oscillation or Pacific Decadal Oscillation. Attribution reveals that warming in the subtropical Northeast Pacific is an important driver of the recently weakened feedback, confirmed by targeted experiments in E3SMv2. Our approach enables near real-time monitoring of Earth's climate stability. %
\\\\%
\textnormal{\textit{Plain Language Summary:} 
As Earth warms, it radiates excess heat to space. This cooling mechanism, known as radiative feedback, keeps the climate stable. The strength of the feedback depends on the geographical pattern of surface temperature changes. We use machine learning and climate model simulations with observed surface temperature reconstructions to estimate how the feedback has changed since the beginning of the 20th century to present day. We find that the climate was most stable in the 1990s. Since then, the feedback has weakened, that is, the climate is currently less efficient at losing heat than in the 1980s and 1990s. We attribute this weakening to spatially changing temperature patterns, especially in the subtropical Northeast Pacific Ocean. By using machine learning to process current temperature fields, we can now monitor changes in Earth's climate stability in near real-time, helping us better understand how the planet's warming rate is evolving.%
\\\\%
\textit{Key points:}\\
$\bullet$\kern0.5em Neural network applied to observed temperature and extended model simulations predict variations of the global feedback parameter up to 2025\\
$\bullet$\kern0.5em Radiative feedbacks experienced a minimum around 1983-2012 but have since been trending towards a less stable climate\\
$\bullet$\kern0.5em The warm subtropical Northeast Pacific weakened the radiative feedback recently\\
}}

\begin{document}

\maketitle

\section{Introduction}

Variations in the Earth's surface temperature influence the energy flux at the top of the atmosphere. The sensitivity of the global-mean radiative response ($R$) to changes in global-mean surface temperature ($\Delta T$) is characterized by the global radiative feedback parameter $\lambda$: $R = \lambda \Delta T$. The global feedback is negative in the current climate, indicating that the Earth is able to release excess energy by warming. A strong (in magnitude) $\lambda$ indicates a more stable climate system, while a weak $\lambda$ indicates a less stable climate system. The Earth's energy imbalance ($N$) is influenced by both $R$ and effective radiative forcing ($F$), such that $N = F + R + \varepsilon = F + \lambda \Delta T + \varepsilon$, with $\varepsilon$ denoting internal variability.

The feedback $\lambda$ varies on decadal timescales due to changing spatial surface temperature patterns \cite[e.g.,][]{Senior00,Armour13,Gregory16,Andrews15,Andrews22,Rugenstein23}. This phenomenon is termed the ``pattern effect''. Variations in $\lambda$ have been linked to lapse rate and low cloud feedbacks, which are sensitive to surface temperature patterns \cite[e.g.,][]{AndrewsWebb18,Ceppi17,Dong19,Zhou16}, but other cloud types, surface albedo (including sea-ice), and water vapor play a role as well \cite[e.g.,][]{Andrews18,Andrews22,Quan25,Zhou25}. 

Since $R$ cannot be observed directly, we often use model simulations to estimate $\lambda$. Coordinated model efforts, such as the Cloud Feedback Model Intercomparison Project \cite[CFMIP3,][]{Webb17}, have allowed for quantification of the pattern effect and its expression in different climate models, by running atmosphere-only climate models with observed sea surface temperature (SST) boundary conditions. However, these simulations end in 2014, limiting our understanding of more recent changes in $\lambda$. Previous estimates have indicated that $\lambda$ weakened (i.e., became less stable) after 2014, coinciding with the strong 2015-2016 El Ni\~{n}o event, a sign change of the Pacific Decadal Oscillation (PDO) index, and enhanced Northeast Pacific warming \citep{Andrews22,Loeb20,Loeb21}.

Here, we use machine learning to estimate $\lambda$ from historical surface temperature reconstructions, allowing us to extend the analysis to the present day \citep{Rugenstein25,VanLoon25}. We use a convolutional neural network (CNN) trained on climate model simulations to learn the relationship between spatial surface temperature and global-mean $R$. We then apply this CNN to different surface temperature reconstructions to estimate historical variations in $R$ and $\lambda$. We compare this statistical approach to extended climate model runs after 2014 and confirm that $\lambda$ has weakened more recently (i.e., more positive feedbacks and a less stable climate system; section~\ref{histfeedback}) and attribute the weaker $\lambda$ to varying surface temperature patterns (section~\ref{attribution}).

\section{Data and methods}\label{methods}

\subsection{Climate model data}\label{methods_data}

We use a combination of simulations from CMIP5 and CMIP6 atmospheric general circulation models (AGCMs) and coupled Earth system models (ESMs), including \emph{historical} \citep{Eyring16}, \emph{piClim-histall} \citep{Pincus16}, and \emph{amip-piForcing} \citep{Webb17} experiments (SI~Table~S1). From each simulation, we use the net top-of-atmosphere radiative flux ($N= {\rm rsdt}-{\rm rsut}-{\rm rlut}$ in CMIP notation, defined as positive downwards) and near-surface (\SI{2}{m}) temperature (tas, ${\bf T}$). We use bold-faced variables to indicate spatial quantities (i.e., on a latitude-longitude grid), while plain variables indicate global values. The \emph{historical} simulations are fully coupled, \emph{piClim-histall} are atmosphere-only runs with preindustrial climatological SST and sea-ice boundary conditions and historically varying forcing agents, and \emph{amip-piForcing} are atmosphere-only runs with historically varying SST and sea-ice boundary conditions and constant forcing at preindustrial levels. 

We use $R$ from \emph{historical} large initial-condition ensemble simulations from one CMIP5 and six CMIP6 ESMs in 1871-2014. All ESMs have at least 20 \emph{historical} ensemble members. We calculate $R_{\rm \emph{hist}} = N_{\rm \emph{hist}} - F_{\rm \emph{piClim-histall}}$, where $N_{\rm \emph{hist}}$ is the net top-of-atmosphere energy flux (from coupled \emph{historical} simulations) and $F_{\rm \emph{piClim-histall}}=N_{\rm \emph{piClim-histall}}$ is the effective radiative forcing from the \emph{piClim-histall} experiment (averaged over all ensemble members if multiple are available). The \emph{historical} near-surface temperature ${\bf T}_{\rm \emph{hist}}$ is taken from the same coupled simulations.

In \emph{amip-piForcing} simulations, $R_{\rm \emph{amip-piForcing}} = N_{\rm \emph{amip-piForcing}}$ since the effective radiative forcing is zero by design. We use \emph{amip-piForcing} simulations from eight CMIP6-generation AGCMs that contributed to CFMIP3 (CFMIP hereafter) to obtain $R_{\rm CFMIP}$ and ${\bf T}_{\rm CFMIP}$ in 1870-2014. We compare the CFMIP ensemble with two extended AGCM experiments. The first (${\bf T}_{\rm HadGEM3},\,R_{\rm HadGEM3}$) is a single ensemble member from 1871-2020 in HadGEM3-GC31-LL \citep{Kuhlbrodt18} with PCMDI AMIP SST boundary conditions \citep{Taylor00} and preindustrial emissions. The second (${\bf T}_{\rm E3SMv2},\,R_{\rm E3SMv2}$) is a set of AGCM experiments performed with the E3SMv2 atmospheric model \cite[][]{Golaz22} and emissions fixed at year-2010 level. The simulations use SST boundary conditions from four reconstructions: HadISST-1.1 \citep{Rayner03}, NOAA ERSST v5 \citep{Huang17}, NOAA OISST v2.1 \citep{Huang21}, and PCMDI AMIP SST \citep{Taylor00}. We use ten ensemble members for each SST dataset spanning 1979-2022. 

We use yearly-mean values as anomalies with respect to 1980-1999. All ${\bf T}$ are bilinearly regridded to a common $\ang{2.8}\times\ang{2.8}$ grid, $R$ is a global mean, and $\Delta T$ is the global-mean \SI{2}{m} temperature anomaly calculated from the regridded dataset. 

\subsection{Convolutional neural network}\label{methods_CNN}

We follow the methodology of \cite{Rugenstein25} and \cite{VanLoon25} to predict the global radiative response $R$ from surface temperature anomalies with a CNN (see SI for details). Whereas \cite{Rugenstein25} and \cite{VanLoon25} used only internal variability to train the CNN, here we include the forced signal. The input to the CNN is ${\bf T}$ and the output is $R$ in the same year. We first train the CNN on ${\bf T}_{\rm \emph{hist}}$ to predict $R_{\rm \emph{hist}}$ from large initial-condition ensemble ESM runs (SI~Table~S1). We use 14 ensemble members from each ESM to train, three for validation, and three for testing. The gradient of the CNN, $\nabla_{\bf T} R_{\rm CNN}$, quantifies the sensitivity of $R$ to temperature perturbations in ${\bf T}$ and can be interpreted as local feedback \cite[similar to, e.g., Green's functions;][]{Zhou17,Dong19,Bloch-Johnson24}. The gradient (SI~Fig.~S2) shows that the pattern of local feedback is consistent with our theoretical understanding of the pattern effect \citep{Rugenstein25}. Moreover, the CNN performs well on unseen test data (SI~Fig.~S3), indicating that the CNN has learned a robust, physical relationship between ${\bf T}$ and $R$.

Then, we finetune the CNN on ${\bf T}_{\rm CFMIP}$ and $R_{\rm CFMIP}$. We take the CNN trained on \emph{historical} data and continue training on \emph{amip-piForcing} data only. Since CNNs generally require large datasets to train, this two-step process allows the CNN to learn from the limited CFMIP dataset. We randomly select 70 years to train, 35 to validate, and 40 to test, and use those years from every AGCM. Validation data is used to stop the training process, to avoid overfitting. Note that, because CFMIP experiments use a historical SST reconstruction \cite[PCMDI AMIP, the same dataset for each AGCM;][]{Taylor00}, the CNN is finetuned on \SI{2}{m} temperature that is strongly constrained by observations over ocean areas. Temperature over land is freely evolving in AGCM experiments. No extended AGCM runs are used to train the CNN. The gradient of the CNN does not change notably after finetuning (SI~Fig.~S2) and hence, historical variability of $\lambda$ does not change much (SI~Figs.~S4-S5). However, finetuning the CNN improves the magnitude of predicted $\lambda$ (Section~\ref{methods_feedback}).

Finally, we apply the trained and finetuned CNN to ${\bf T}_{\rm rc}$ from different spatially complete reconstructions (rc; SI~Table~S2). We use \SI{2}{m} temperature from three atmospheric reanalyses [\cite[ERA5,][]{Hersbach20}; \cite[JRA-3Q,][]{Kosaka24}; and \cite[MERRA-2,][]{Gelaro2017}] and four datasets that combine SST with land surface air temperature [\cite[Berkeley Earth,][]{Rohde20}; \cite[DCENT-I,][]{Chan26}; \cite[HadCRUT5,][]{Morice21}; and \cite[NOAAGlobalTempV6,][]{NOAAGlobalTemp}]. Spatial SST anomalies are similar to \SI{2}{m} temperature anomalies. We process ${\bf T}_{\rm rc}$ the same as the climate model data: yearly anomalies are defined with respect to 1980-1999 and bilinearly regridded to $\ang{2.8}\times\ang{2.8}$. Reconstructions ${\bf T}_{\rm rc}$ are not used to train the CNN, but only for inference.

We use SHapley Additive exPlanations (SHAP) values \citep{Lundberg17} to decompose $R_{\rm CNN}$ into contributions from each grid box. SHAP is an eXplainable AI attribution technique. SHAP values ${\bf S}$ constitute an attribution map that quantifies how much the input ${\bf T}$ contributes to the output $R_{\rm CNN}$. SHAP values are complete, such that $R_{\rm CNN} = \sum_j S_j$, with $j$ denoting the grid box.

\subsection{Differential global feedback parameter} \label{methods_feedback}

\begin{figure*}
    \centering
    \includegraphics[width=\textwidth]{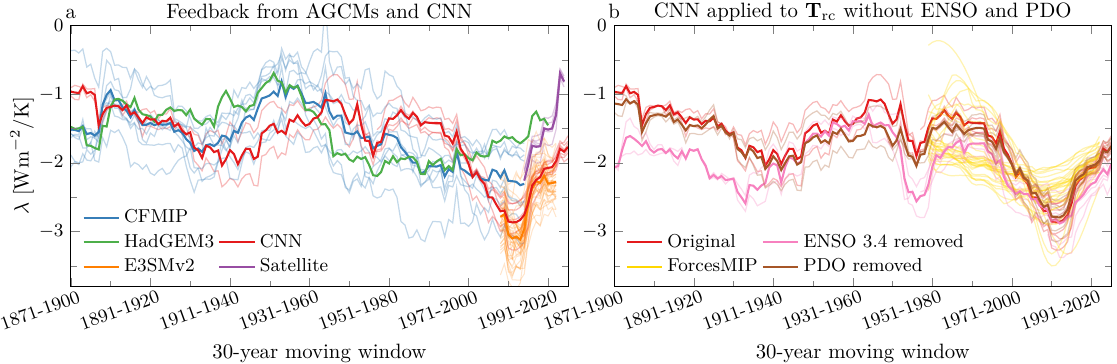}
    \caption{
        (a) The global feedback parameter $\lambda$ estimated from AGCMs that contributed \emph{amip-piForcing} experiments to CMIP6 (CFMIP, blue, ending in 2014), an extended \emph{amip-piForcing} experiment in HadGEM3-GC31-LL (green, ending in 2020), AMIP-F2010 experiments in E3SMv2 (orange, ending in 2022), the CNN applied to seven historical surface temperature reconstructions (red, ending in 2025), and an estimate based on satellite observations (satellite, purple, ending in 2024). 
        (b) The global radiative feedback parameter $\lambda$ estimated from the CNN applied to ${\bf T}_{\rm rc}$ (red, as in a); ${\bf T}^{{\neg I_{3.4}}}_{\rm rc}$ with the El Ni\~{n}o Southern Oscillation (ENSO) linearly removed (as defined by the Ni\~{n}o 3.4 region; pink); ${\bf T}^{{\rm \neg PDO}}_{\rm rc}$ with the Pacific Decadal Oscillation (PDO) linearly removed (brown).
        Thin lines are individual model simulations (for CFMIP), ensemble members (for E3SMv2), or temperature reconstructions (for CNN and satellite); thick lines are a mean across the thin lines per dataset.
    }
    \label{historical_feedback}
\end{figure*}

We use the differential feedback parameter to study temporal variations of $\lambda$ \citep{Andrews15,Andrews22,Rugenstein21}. In a given 30-year period ending in year $t$, we estimate $\lambda(t) = \left.\partial R/\partial\Delta T\right\vert_{[t-29,t]}$ as the linear regression slope of $\{R(t-29),\dots,R(t)\}$ against global-mean near-surface temperature $\{\Delta T(t-29),\dots,\Delta T(t)\}$. The feedback $\lambda(t)$ thus represents how $R$ linearly varies with $\Delta T$ over moving 30-year windows and their associated surface temperature patterns. Each $\lambda(t)$ and temperature pattern can be the result of internal variability, external forcing, or a combination of both \cite[e.g.][]{Rugenstein21,Lin25}. We calculate $\lambda$ directly from AGCM output ($\lambda_{\rm CFMIP}$, $\lambda_{\rm HadGEM3}$, $\lambda_{\rm E3SMv2}$) and from the CNN applied to reconstructions ($\lambda_{\rm CNN}$).

We compare $\lambda_{\rm CNN}$ to another observation-based estimate. We use $N$ from a reconstruction \cite[DEEP-C,][]{Allan14,Liu20} and observations \cite[CERES-EBAF,][]{Loeb18} combined with the historical forcing estimate from \cite{Forster25} to calculate $R_{\textnormal{satellite}} = N_{\textnormal{satellite}} - F_{\textnormal{Forster}}$ and then $\lambda_{\textnormal{satellite}} = \partial R_{\textnormal{satellite}}/\partial\Delta T_{\rm rc}$, using different reconstructions (SI~Table~S2) for global mean temperature $\Delta T_{\rm rc}$. Because $N_{\textnormal{satellite}}$ uses different satellite products throughout the record (none of which are 30 years long), we cannot exclude that changes in $\lambda_{\textnormal{satellite}}$ are due to inconsistencies between observational products, but we can qualitatively use $\lambda_{\textnormal{satellite}}$ as an independent line of evidence for recent variations in $\lambda$.

SHAP values are used to attribute $\lambda$ to surface temperature patterns. We calculate the local contribution to $\lambda(t)$ by regressing the SHAP values in each grid box $j$ to global mean temperature to obtain $L_j(t) = \left.\partial S_j/\partial\Delta T\right\vert_{[t-29,t]}$, with $\lambda(t) = \sum_j L_j(t)$. Similarly, we define the surface temperature pattern $P_j(t) = \left.\partial T_j/\partial\Delta T\right\vert_{[t-29,t]}$. In a given 30-year period, ${\bf P}(t)$ indicates how surface temperature varies with global mean temperature, while ${\bf L}(t)$ indicates the contribution of ${\bf P}(t)$ in each grid box to the global feedback $\lambda(t) = \sum_j L_j(t)$.

\section{Historical variations of the global radiative feedback parameter} \label{histfeedback}

We confirm that $\lambda(t)$ varied substantially throughout the twentieth century (Fig.~\ref{historical_feedback}a). Estimates from the CNN and AGCM experiments show a maximum in the early-to-mid twentieth century ($\lambda\simeq\SI{-1}{Wm^{-2}/K}$) and a minimum in the late twentieth to early twenty-first century, although the exact maximum and minimum periods differ. Some AGCMs show more pronounced decadal variability than others (SI~Fig.~S6). HadGEM3-GC31-LL shows a more gradual variation compared to the CNN and E3SMv2 estimates. The latter two indicate a minimum around 1981-2010 to 1985-2014 ($\lambda\simeq\SI{-3}{Wm^{-2}/K}$) and a sharp increase (weakening) afterwards ($\lambda\simeq\SI{-2}{Wm^{-2}/K}$). The E3SMv2 ensemble mean predicts $\lambda \lesssim \SI{-3}{Wm^{-2}/K}$ at its minimum, which is more negative than most other estimates, suggesting that E3SMv2 is particularly sensitive to varying SST patterns \citep{Qin22,Qin24}.

The sharp increase in $\lambda$ after 1985-2014 is confirmed by the satellite-based estimate (purple lines in Fig.~\ref{historical_feedback}), in line with \cite{Loeb20} and \cite{Andrews22}. However, $\lambda_{\rm satellite}$ is weaker than $\lambda_{\rm CNN}$ and $\lambda_{\rm E3SMv2}$, reaching $\lambda_{\rm satellite} \gtrsim \SI{-1}{Wm^{-2}/K}$ in 1995-2024. This might be due to the estimate of effective radiative forcing from \cite{Forster25} used to calculate $R_{\rm satellite}$ and thus $\lambda_{\rm satellite}$ (Section~\ref{methods_feedback}). \cite{VanLoon25} suggest that the \cite{Forster25} forcing trend estimate is biased low in the last two decades, implying an unphysical underestimation of $\lambda$. The CNN estimate $\lambda_{\rm CNN}$ uses the same method as \cite{VanLoon25} to determine $R$ and is thus consistent with the strong forcing trend found therein.

The mean CNN estimate deviates from the CFMIP ensemble mean, even though the CNN was finetuned on that dataset. When applying the CNN to ${\bf T}_{\rm CFMIP}$, the CNN predicts $\lambda$ similar to the CFMIP multimodel mean (SI~Fig.~S4). Therefore, $\lambda_{\rm CNN}$ has a ``model-average'' pattern effect, because it has been trained on AGCMs with a wide range of temporal $\lambda$ variability (SI~Fig.~S6). The CNN could be finetuned on separate models to investigate model differences in the pattern effect, which we leave for future work. The difference between $\lambda_{\rm CNN}$ and $\lambda_{\rm CFMIP}$ is mainly explained by temperature over land, but differences in near-surface temperature over ocean also play a role, even though SST is the same in each CFMIP simulation. Reconstructions ${\bf T}_{\rm rc}$ include observed land surface temperature, while ${\bf T}_{\rm CFMIP}$ has simulated land temperature. Land surface temperature in ${\bf T}_{\rm CFMIP}$ neglects direct warming due to forcing agents (e.g., CO$_2$ emissions), that are not included in \emph{amip-piForcing} experiments \cite[e.g.,][]{Andrews14,Andrews22}.

\subsection{Spread in historical feedback estimates}

The large spread in $\lambda$ estimates (Fig.~\ref{historical_feedback}, thin lines) can be due to differences in AGCMs, internal atmospheric variability, or near-surface temperature, both over ocean and land. Averaged over all years, the intermodel standard deviation is $\sigma_{\rm CFMIP} \simeq \SI{0.43}{Wm^{-2}/K}$ across CFMIP AGCMs (Fig.~\ref{historical_feedback}, thin blue lines). Similarly, the standard deviation across E3SMv2 simulations with different SST reconstructions is $\sigma_{\rm E3SMv2} \simeq \SI{0.30}{Wm^{-2}/K}$ (Fig.~\ref{historical_feedback}, thin orange lines). Since $\sigma_{\rm E3SMv2}$ does not include model differences, the variance explained due to model differences is $r^2_{\rm model} = 1 - \sigma_{\rm E3SMv2}^2/\sigma_{\rm CFMIP}^2 \simeq \SI{50}{\percent}$. The standard deviation $\sigma_{\rm CNN}$ does not change significantly when applying the CNN to ${\bf T}$ output from CFMIP or E3SMv2 instead of reconstructions (SI~Fig.~S4). Because $\sigma_{\rm CNN}$ only includes differences in ${\bf T}$, the variance explained due to differences in models plus internal atmospheric variability is $r^2_{\rm model+IV} = 1 - \sigma_{\rm CNN}^2/\sigma_{\rm CFMIP}^2 \simeq \SI{80}{\percent}$. In other words, about $r^2_{\bf T} \simeq \SI{20}{\percent}$ of the $\lambda_{\rm CFMIP}$ variability in a given year can be explained by differences in near-surface temperature reconstructions alone.

AGCM experiments do not prescribe land temperature, which is freely evolving in response to the prescribed SST. Since CFMIP simulations all use the same SST boundary conditions, we expect most of $r^2_{\bf T} \simeq \SI{20}{\percent}$ to be explained by land temperature variability. To test this, we create an artificial temperature dataset ${\bf T}_{\rm CFMIP,\,land}$ where we replace near-surface temperature over ocean in ${\bf T}_{\rm CFMIP}$ with the CFMIP multimodel mean, while keeping land temperature unchanged. We then apply the CNN to ${\bf T}_{\rm CFMIP,\,land}$ and calculate $\lambda$. We find that about $r^2_{\rm land\,{\bf T}} \simeq \SI{85}{\percent}$ of the variance due to surface temperature differences is explained by land temperature. We find similar results using the E3SMv2 ensemble. When performing the same analysis on ${\bf T}_{\rm rc}$, we find instead $r^2_{\rm land\,{\bf T}}\simeq\SI{45}{\percent}$, because ocean temperature differences across reconstructions are larger than across CFMIP simulations.

In summary, half of the intermodel spread of $\lambda$ can be explained by AGCM differences ($r^2_{\rm model} \simeq \SI{50}{\percent}$), while differences in near-surface temperature explain $r^2_{\bf T} \simeq \SI{20}{\percent}$ (of which $r^2_{\rm land\,{\bf T}} \simeq \SI{85}{\percent}$ is due to temperature over land). The remaining variance is explained by internal atmospheric variability. The sensitivity of $\lambda$ to observational uncertainties \cite[e.g.,][]{Lewis21,Andrews22,Modak23,Fan25} can be investigated by applying the CNN to different reconstructions (SI~Fig.~S7). Using the 200-member HadCRUT5 ensemble \citep{Morice21}, we find that reconstruction uncertainties lead to a spread of about $\pm\SI{0.24}{Wm^{-2}/K}$ (\SI{99}{\percent} confidence interval) in the satellite era. Using output from the E3SMv2 ensemble, we find a similar spread of about $\pm\SI{0.28}{Wm^{-2}/K}$. Future work could investigate these uncertainties in more detail, by comparing attribution maps of the CNN applied to different reconstructions or using AGCMs forced with different reconstructions (e.g., the E3SMv2 ensemble).

\subsection{Role of the El Ni\~{n}o Southern Oscillation and Pacific Decadal Oscillation}

It is an open question to what extent the historical variations in $\lambda$ are forced or due to internal variability \cite[e.g.,][]{Proistosescu18,Lutsko18,Lin25}. El Ni\~{n}o Southern Oscillation (ENSO) and Pacific Decadal Oscillation (PDO) have their own pattern effect \cite[e.g.,][]{Loeb20,Ceppi21,Wills21,Meyssignac23,Guillaume-Castel25}. We remove the influence of ENSO and PDO on our $\lambda$ estimate. First, we regress ${\bf T}_{\rm rc}$ at each grid box onto the Ni\~n{o} 3.4 index $I_{3.4}$ on monthly timescales to obtain the part of ${\bf T}_{\rm rc}$ that is linearly related to $I_{3.4}$, denoted ${\bf T}^{{I_{3.4}}}_{\rm rc}$. Then, we subtract the $I_{3.4}$-related variability and take yearly averages, to create an artificial timeseries ${\bf T}^{{\neg I_{3.4}}}_{\rm rc} = {\bf T}_{\rm rc} - {\bf T}^{{I_{3.4}}}_{\rm rc}$ that excludes effects of ENSO. Detrending $I_{3.4}$ does not significantly affect the regression because monthly variability is much larger than decadal trends. Using other methods to remove ENSO \cite[e.g.,][]{Gunnarson25} or simply leaving out strong ENSO years does not alter our conclusions (SI~Fig.~S8a). We do the same for the monthly PDO index \citep{Zhang97,NOAA_NCEI_PDO} to create ${\bf T}^{{\rm \neg PDO}}_{\rm rc} = {\bf T}_{\rm rc} - {\bf T}^{{\rm PDO}}_{\rm rc}$. Finally, we apply the CNN to ${\bf T}^{{\neg I_{3.4}}}_{\rm rc}$ and ${\bf T}^{{\rm \neg PDO}}_{\rm rc}$ to investigate the influence of ENSO and PDO on $\lambda$ (Fig.~\ref{historical_feedback}b). This procedure does not remove long-term El-Ni\~{n}o-like or PDO-like trends or lagged effects, but only the variability that is linearly related to the respective indices on monthly timescales.

Removing ENSO generally strengthens the feedback, indicating that ENSO tends to weaken $\lambda$ in the 30-year regression definition used here. The same is true for PDO, although the effect is small. The effect of ENSO is largest in early periods, when internal variability is much larger than decadal trends. Different ENSO indices can alter $\lambda$ to various degrees, with the Ni\~{n}o 1+2 region weakening the feedback the most, possibly because of its proximity to low cloud decks (SI~Fig.~S8b). ENSO can have an effect on $R$ and $\Delta T$ in a given year \cite[e.g.,][]{Ceppi21,Guillaume-Castel25}, but only weakly influences how $R$ varies with $\Delta T$ in a 30-year period (SI~Fig.~S8c-e). The weak influence of ENSO and PDO on the $\lambda$-minimum around 1981-2010 indicates that there is likely a forced component to the pattern effect on decadal timescales. This is confirmed by applying the CNN to estimates of the forced component of observed ${\bf T}$ \cite[ForceSMIP,][who compared 30 different statistical methods to separate the forced signal from interval variability]{ForceSMIP}. Although there is no consensus on the forced signal in observed ${\bf T}$, most estimates indicate a $\lambda$-minimum around 1981-2010 (Fig.~\ref{historical_feedback}b, yellow). Since the CNN can make fast predictions for any temperature pattern, it can be used to test such hypotheses by applying it to artificial surface temperature anomalies.

\section{Attributing changes in the global radiative feedback parameter to surface temperature patterns} \label{attribution}

\begin{figure*}
    \centering
    \includegraphics{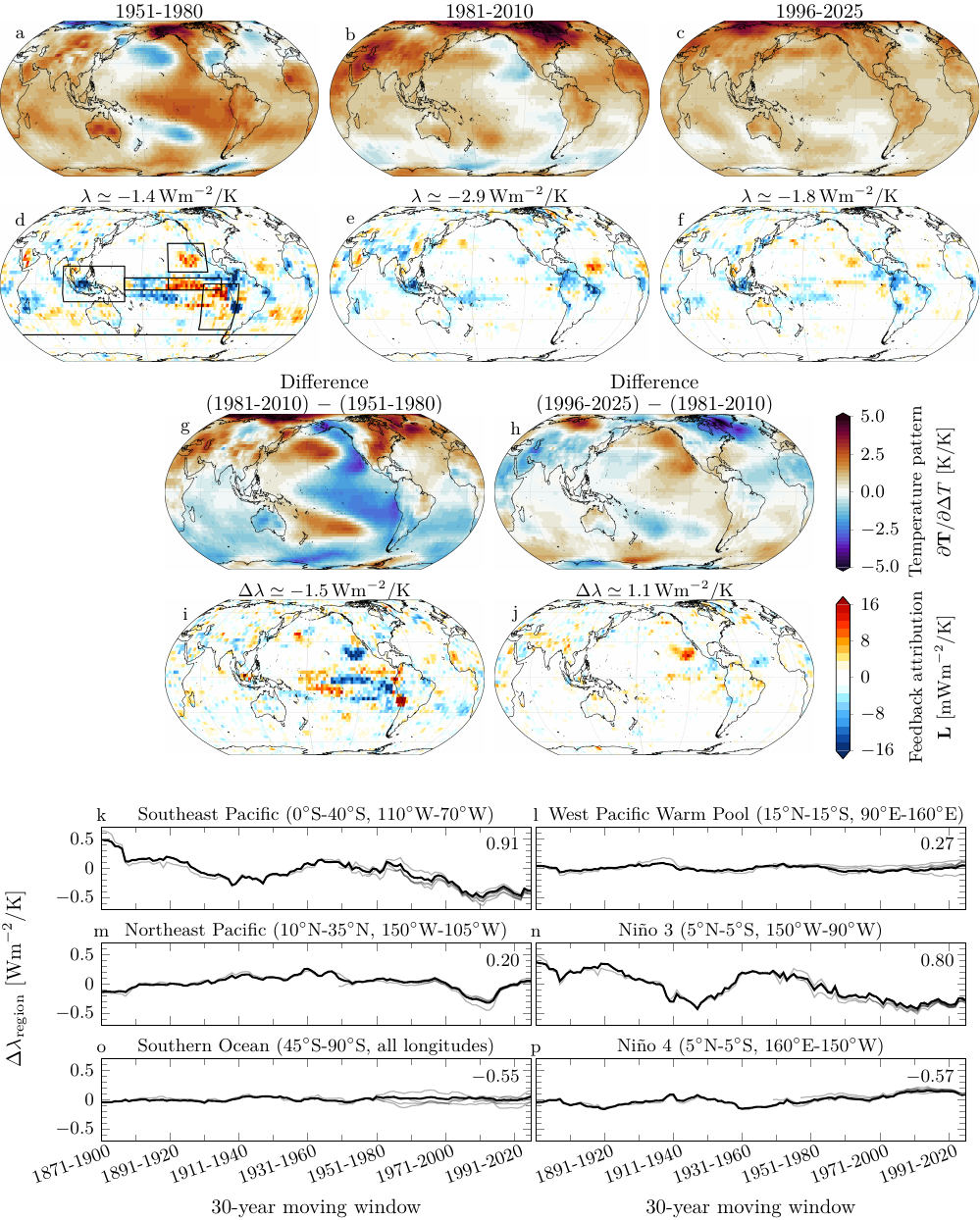}
    \caption{
        Attribution of changes in the global radiative feedback parameter. Panels a-c show the surface temperature patterns (linear regression slope of local temperature against global mean temperature). Panels d-f show the local temperature contribution to the feedback parameter. The sum over all grid boxes equals $\lambda$ in the same period (indicated above the maps). Bottom panels show the difference between two periods for the surface temperature pattern (g-h) and feedback attribution maps (i-j). Maps are averaged over seven reconstructions ${\bf T}_{rc}$, except for 1951-1980, with only 3 available datasets. All temperature patterns and attribution maps share their respective colorbars. Panels k-p show the contribution of six different regions (see panel a) to the global feedback parameter, considering only ocean area in each box. All values are anomalies with respect to the 1871-2025 mean. Thin lines are individual ${\bf T}_{\rm rc}$, thick lines are an ensemble mean. Values on the top right indicate the Pearson correlation coefficient with the global feedback (Fig.~\ref{historical_feedback}).
    }
    \label{attr}
\end{figure*}

Which spatial patterns and physical processes led to the historical variations of $\lambda$, in particular the minimum around 1983-2012? We attribute the change in $\lambda$ to changes in surface temperature patterns using SHAP values (explained in section~\ref{methods}). Fig.~\ref{attr} shows surface temperature patterns and feedback attribution maps for three different periods: 1951-1980, 1981-2010, and 1996-2025 (note an overlap between the latter two periods). The temperature patterns indicate how local surface temperature varies with global mean temperature in each period and are not necessarily indicative of regional cooling or warming in that period. The feedback attribution maps show how much temperature in each grid box contributes to the global feedback parameter, through both local and remote processes. The attribution maps are noisy, but the large-scale patterns are robust across different CNN architectures or training data.

The year 1980 is considered a turning point for the pattern effect \citep{Fueglistaler21,Andrews22,Wu25}, with $\lambda$ strengthening (stabilizing) for 30-year windows ending after 1980 (Fig.~\ref{historical_feedback}). After 1980, surface temperature patterns changed notably in the Pacific Ocean, with slower warming or even cooling in the eastern Pacific (Fig.~\ref{attr}a,b,g). This pattern change is associated with a weakened positive feedback contribution from the equatorial and subtropical eastern Pacific (Fig.~\ref{attr}d,e,i). This is consistent with a switch from positive to negative low cloud feedbacks over the same period, driven mostly by marine cloud decks in the subtropical eastern Pacific \citep{Myers23}. Other regions contributed positively to the feedback, such as the central Pacific and the West coast of South America, but the global net effect was a stronger, more stabilizing feedback ($\Delta\lambda \simeq \SI{-1.6}{Wm^{-2}/K}$ between 1951-1980 and 1981-2010). 
The strong feedbacks in the early 2000s might be connected to stabilized global mean temperatures during the global warming hiatus \cite[e.g.,][]{Medhaug17,Hedemann17,Modak21}, but shorter timescales than 30 years need to be considered since the hiatus lasted only around 2000-2012.

For 30-year windows ending after 2010, $\lambda$ rapidly weakened (Fig.~\ref{historical_feedback}). This weakening is associated with a sign-change of the temperature pattern in the Northeast Pacific (Fig.~\ref{attr}b,c,h), which contributed to a strong positive feedback (Fig.~\ref{attr}f,j). Also the equatorial East Pacific contributed positively to the feedback. These positive contributions appear once 2015-2016 is included in the 30-year window, consistent with prior studies \citep{Andrews22,Loeb20}, and intensify over subsequent years. Removing 2015-2016 from the regression does not significantly affect the increased $\lambda$, indicating that the weakening trend is not due to a single ENSO event. The subtropical North Atlantic contributed less positively to the feedback after 1981-2010, but globally the feedback weakened ($\Delta\lambda \simeq \SI{1.1}{Wm^{-2}/K}$ between 1981-2010 and 1996-2025). 
In 1996-2025, $\lambda$ is still stronger than predicted by abrupt CO$_2$ forcing experiments \cite[e.g.,][]{Andrews22}, possibly due to residual effects of other climate forcings \cite[e.g.,]{Gregory20}. Delayed warming in the Southeast Pacific kept $\lambda$ stronger relative to uniform warming (SI~Fig.~S9). With more uniform warming, $\lambda$ would weaken further, getting closer to $\lambda$ predicted from CO$_2$-only forcing experiments.

Over the full historical period, different regions contributed to temporal variability of $\lambda$. Ocean temperature dominates the variability of $\lambda$, but land temperature has a considerable contribution. The CNN attributes the overall magnitude of the feedback mostly to land temperature changes (SI~Fig.~S10), possibly because of the stronger global warming signal over land. Therefore, we do not use the CNN to attribute the overall magnitude of $\lambda$, but only its predicted variations in time. Figs.~\ref{attr}k-p show the regional attribution of changes in $\lambda$, by summing attribution maps over regions indicated in Fig.~\ref{attr}d. 

Among different ocean regions, the Southeast Pacific (Fig.~\ref{attr}k) and Ni\~{n}o~3 (Fig.~\ref{attr}n) regions correlate most closely with the global feedback and contributed to the strengthening after 1951-1980. The contribution from the Ni\~{n}o~4 region anticorrelates with $\lambda$ (Fig.~\ref{attr}p), in line with results of \cite{Guillaume-Castel25}, who found that different ENSO flavors can influence top-of-atmosphere radiation in opposite ways on inter-annual timescales. 
The high correlation of $\lambda$ with its contribution from ENSO regions indicate that long-term variability in the Tropical Pacific has a strong influence on the pattern effect, even though the influence of single ENSO events on $\lambda$ variations is relatively weak (Fig.~\ref{historical_feedback}b and SI~Fig.~S8), highlighting the distinction between decadal trends and interannual variability.
After 1981-2010, the Northeast Pacific (NEP, Fig.~\ref{attr}m) stands out as a contributor to the weakening of $\lambda$ (see also Fig.~\ref{attr}j). Between 1981-2010 and 1996-2025, the NEP contributed $\Delta\lambda_{\rm NEP} \simeq \SI{0.33}{Wm^{-2}/K}$ to the feedback weakening ($\sim\SI{30}{\percent}$ of the total change). The West Pacific Warm Pool (WPWP, Fig.~\ref{attr}l) and Southern Ocean (Fig.~\ref{attr}o) have the weakest contribution to the global feedback variations. The WPWP contributes to the overall negative sign of the feedback (Fig.~\ref{attr}d-f), but the temperature has little interannual to decadal variability (Fig.~\ref{attr}a-c, g-h). Since the relative warm pool temperature with respect to the tropical-mean temperature is more important for the pattern effect than the absolute temperature \cite[e.g.,][]{Dong19,Fueglistaler21}, the CNN might attribute some changes in the feedback to where the temperature and top-of-atmosphere radiation change are occurring, not where they are caused (e.g., low clouds in the East Pacific might respond to warming in the WPWP). The CNN in its current setup cannot distinguish between these local versus remote processes.

\begin{figure*}
    \centering
    \includegraphics[width=\textwidth]{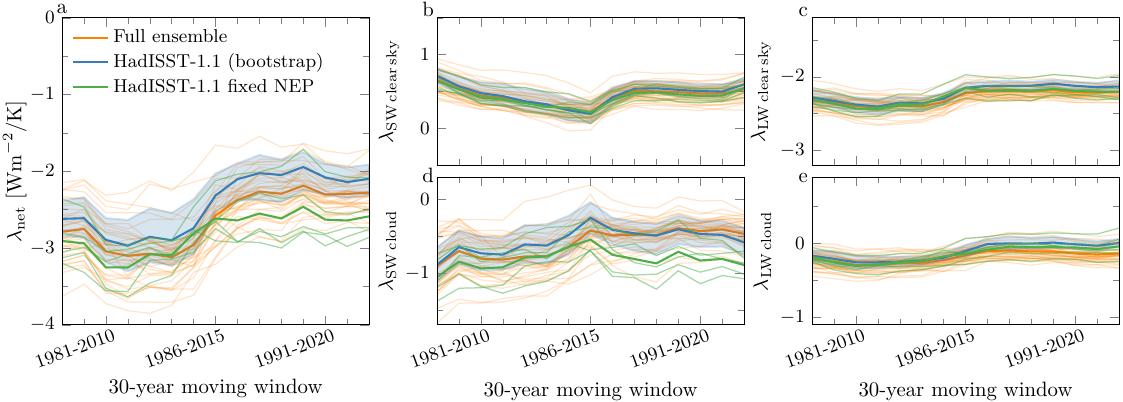}
    \caption{
        Decomposition of the net global feedback (a) into its shortwave (SW) clear-sky (b), longwave (LW) clear-sky (c), SW cloud (d), and LW cloud (e) components in the E3SMv2 ensemble (orange). Five members of E3SMv2 are run with HadISST-1.1 boundary conditions and fixed Northeast Pacific (NEP) SSTs (green). Thin lines are individual ensemble members, thick lines are an ensemble mean. We compare with bootstrapped 5-member averages from the 10-member E3SMv2 ensemble with HadISST-1.1 boundary conditions. Thick blue line shows the median, shading the two-sided \SI{95}{\percent} confidence bounds.
    }
    \label{type}
\end{figure*}

The net radiative feedback is influenced by many different processes, including lapse rate, cloud, and sea-ice feedbacks. Decomposing the net radiation into short-wave/long-wave and clear-sky/cloud components can give additional insights into these different components \citep{Andrews18}. Cloud fluxes are calculated as all-sky minus clear-sky fluxes. Training CNNs on the the different components is beyond the scope of this paper, but we investigate the decomposition in the E3SMv2 ensemble and find that all components of $\lambda$ weaken after 1985-2014 (Fig.~\ref{type}). Aside from variations in the shortwave cloud feedback (predominantly driven by low clouds) and longwave clear-sky feedback (predominantly lapse rate feedback), also the shortwave clear-sky feedback shows a significant increase after 1986-2015. The latter might be due to a decline in Antarctic sea-ice after 2015 \cite[e.g.,][]{Riihela21,Duspayev24,Zhou25}. The local feedback in the Southern Ocean shows a positive change of the same order as the feedback change in the NEP (SI~Fig.~S11), mainly driven by the SW clear-sky feedback, highlighting the role of Antarctic sea-ice in the weakening of $\lambda$.

We confirm the relevance of the NEP in E3SMv2. We perform five additional AGCM simulations in E3SMv2 with HadISST-1.1 boundary conditions and year-2010 emissions, but keep SST in the NEP fixed to 1979 values. Keeping the NEP SST fixed reduces the overall feedback, mostly due to the SW cloud component (Fig.~\ref{type}), since the NEP is dominated by low cloud feedbacks. We compare the fixed-NEP to the original 10-member ensemble of E3SMv2 with HadISST-1.1 boundary conditions (Section~\ref{methods_data}). We sample five members from all 10 members with replacement, take the 5-member average, and repeat this process 1000 times to obtain the median 5-member mean (Fig.~\ref{type}, blue) and the two-sided \SI{95}{\percent} confidence bounds (Fig.~\ref{type}, shading). From 1986-2015 onward, the SW cloud feedback in the fixed-NEP experiment is lower than in the original ensemble (Fig.~\ref{type}d), which results in a lower net feedback in 1993-2022 (Fig.~\ref{type}a). No other feedback components are significantly affected by fixing NEP SST.

\section{Discussion and conclusions}

The CNN is a surrogate model that relates any surface temperature pattern to $R$ and $\lambda$. Therefore, the CNN can be used for fast experimentation on the effect of temperature variations on $\lambda$, exemplified here by linearly removing ENSO and PDO from ${\bf T}_{\rm rc}$ (Fig.~\ref{historical_feedback}b) or local attribution (Fig.~\ref{attr}). Such experiments can inform climate model runs, as done here to study the influence of the Northeast Pacific on the recent $\lambda$ weakening in E3SMv2, to gain robust understanding of the pattern effect. 

We update $\lambda$ with various surface temperature reconstructions up to present day. This means that we can now monitor changes in $\lambda$ in near real-time, without relying on new climate model simulations, currently lagging more than a decade. Our estimate generally follows the CFMIP3 multi-model mean before 2014, but additional uncertainty could be introduced by applying the CNN to much warmer (out-of-distribution) surface temperature maps. Because the CNN uses global surface temperature maps, not just ocean temperature, we capture the effect of land temperature on $\lambda$. In future work, it is important to compare the CNN results with climate model simulations with fixed land surface temperature \cite[e.g.,][]{Andrews21} to confirm the importance of land temperature variability for $\lambda$ variations in a causal framework instead of a statistical one.

Regional attribution of changes in $\lambda$ confirms the importance of the East Pacific for the pattern effect \cite[e.g.,][]{Andrews22,Loeb20,Myers23}. The CNN attributes the strengthening of $\lambda$ after 1980 to cooling in the tropical and subtropical eastern Pacific, while the weakening after 2010 is mainly attributed to warming in the subtropical Northeast Pacific (Fig.~\ref{attr}). AGCM experiments with SSTs fixed in this region confirms the regional attribution suggested by the CNN.

We only use near-surface temperature to predict top-of-atmosphere radiation with the CNN. In reality, other factors, such as sea-ice or direct aerosol effects, also play a role \cite[e.g.,][]{Riihela21,Andrews22,Duspayev24,Zhou25}. Still, the CNN performs well on unseen test data, indicating that near-surface temperature contains most of the information needed to predict $R$, with $r^2$ values above \SI{80}{\percent} (SI~Fig.~S3). Future work could include sea-ice as a predictor to the CNN, which can give additional insight into the SW clear-sky feedback (Fig.~\ref{type}b). AGCM experiments with varying SST but fixed sea-ice could confirm the role of sea-ice on the recent weakening of the feedback.

In summary, the climate was the most stable in recent history between the mid-1980s and early 2010s ($\lambda \simeq \SI{-3}{Wm^{-2}/K}$). Over the last three decades, $\lambda$ has been less stable ($\lambda \simeq \SI{-2}{Wm^{-2}/K}$). This result is found both with a statistical (CNN) approach and in two AGCMs (HadGEM3-GC31-LL and E3SMv2). Temperature pattern differences between 1981-2010 and 1996-2025 are opposite in sign to those between 1951-1980 and 1981-2010 (Fig.~\ref{attr}g-h), suggesting that the climate is returning to a state with a less stabilizing feedback, possibly explaining the recent accelerated warming \citep{Alessi23,Foster26}. If the current surface temperature pattern persists, future global mean warming rates might be large (possibly larger than anticipated by freely running climate models). Weakening of $\lambda$ might be counteracted or enhanced by ocean heat uptake and ${\bf T}$, calling for research on their local drivers and interactions.

%
%

\section*{Open Research Section}
We acknowledge the World Climate Research Programme, which, through its Working Group on Coupled Modeling, coordinated and promoted CMIP6. We thank the climate modeling groups for producing and making available their model output, the Earth System Grid Federation (ESGF) for archiving the data and providing access, and the multiple funding agencies who support CMIP6 and ESGF. All training data used in this manuscript can be accessed freely from https://esgf-node.ornl.gov/ for CMIP model output. Near-surface temperature reconstructions are available through \cite{ERA5DATA,MERRA2DATA,JRA3QDATA,NOAAGlobalTempDATA,DCENTdata,Morice21,Rohde20} and top-of-atmosphere fluxes from \cite{DEEPCdata,CERESEBAF42}. Software and global mean values of $T$, $R$, and $\lambda$ from AGCM and CNN output can be found at \cite{VanLoon26}.

\acknowledgments
TA was supported by the Met Office Hadley Centre Climate Programme funded by DSIT. The work of MDZ was supported by the U.S. Department of Energy (DOE) Regional and Global Model Analysis Program and was performed under the auspices of the US DOE by Lawrence Livermore National Laboratory under contract DEAC52-07NA27344. This research used resources of the National Energy Research Scientific Computing Center (NERSC), a Department of Energy User Facility, using NERSC award BER-ERCAP0033047. SVL and MR were supported, in part, by the National Science Foundation (NSF) under Grants 2233673 and 2530919. SVL and MR thank Elizabeth A. Barnes for discussions. Generative AI was used to assist with software development and text clarity, but all research conceptualization, design, analysis, interpretation, and conclusions are original to the authors, who take full responsibility for the content's accuracy and academic integrity.

\clearpage
\onecolumn
\section*{Supporting Information}
\setcounter{figure}{0}
\setcounter{table}{0}
\renewcommand{\figurename}{{\bf SI Fig.}}
\renewcommand{\tablename}{{\bf SI Table}}
\renewcommand{\thefigure}{{\bf S\arabic{figure}}}
\renewcommand{\thetable}{{\bf S\arabic{table}}}

\noindent\textbf{Contents of this file}
\begin{enumerate}
\item Text S1
\item Figures S1 to S11
\item Tables S1 to S3
\end{enumerate}

\section*{Text S1: Details on the CNN architecture and training procedure}

\begin{figure}[ht]
    \centering
    \includegraphics{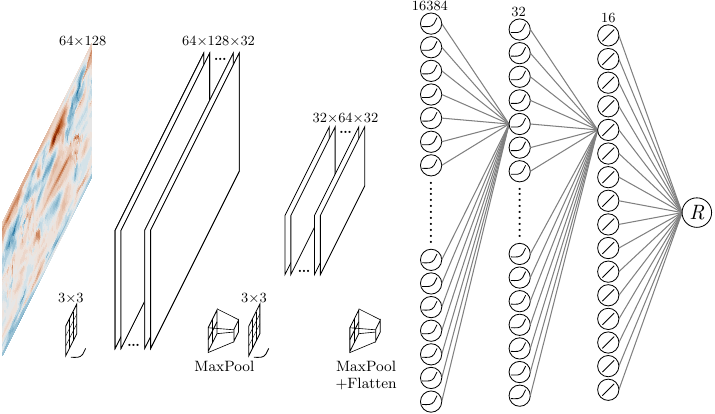}
    \caption{Schematic of the CNN architecture used. The input maps (\SI{2}{m} temperature) are passed through two convolutional layers, each with 32 kernels of size $3\times3$, followed by a max pooling layer and an ELU activation function. The result is flattened and passed through two fully connected layers with 32 and 16 neurons, with a ELU and linear activation function respectively. The final result is a single number estimating the global-mean radiative response $R$. Adapted from \cite{Rugenstein25} and \cite{VanLoon25}.
    }
    \label{CNN_architecture}
\end{figure}

\begin{figure}[ht]
    \centering
    \includegraphics[width=\textwidth]{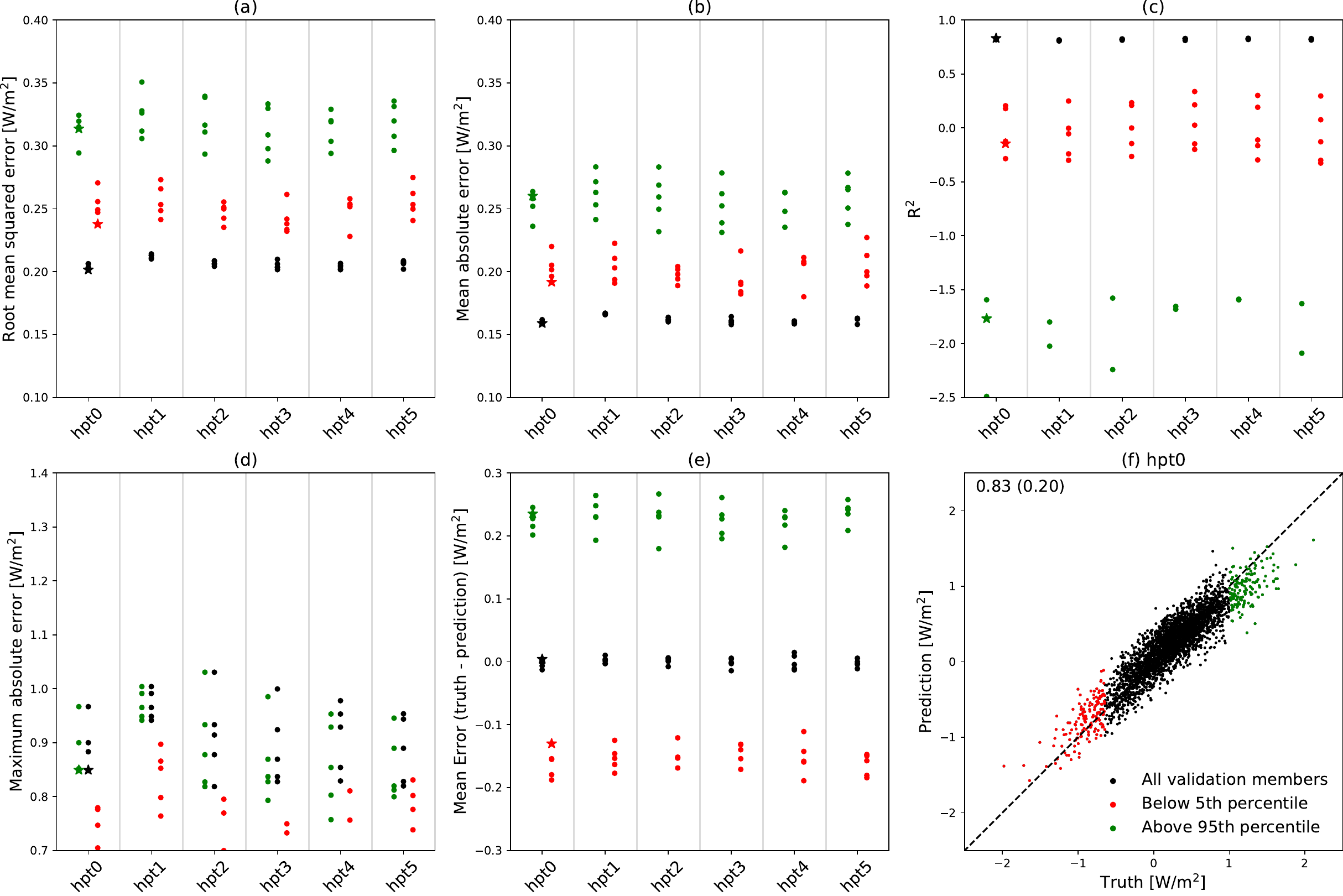}
    \caption{
        Hyperparameter testing for different CNN architectures. The $x$-axis labels are defined in Supporting Information Tab.~\ref{tab:hpts}, and each dot represents a different random initialization of the CNN before training. All results shown in this plot are for the initial training only (i.e., using \emph{historical} data from coupled models). For each trained CNN, we compute the (a) root mean squared error, (b) mean absolute error, (c) $R^2$ value, (d) maximum absolute error, and (e) mean error (truth$-$prediction). Panel (f) reports the true $R$ versus predicted $R$ of all years in the validation dataset. Black dots represent all years, while red and green dots show only those in the lower and upper 5th percentiles, to examine how well the CNN performs on the extremes. The CNN used in the main text is highlighted by a star. Adapted from \cite{Rugenstein25} and \cite{VanLoon25}.
    }
    \label{hpts}
\end{figure}

We use the same CNN architecture as in \cite{VanLoon25}, see Fig.~\ref{CNN_architecture} for a schematic. This architecture was chosen after trying multiple hyperparameter setups, as shown in Fig.~\ref{hpts}. Hyperparameter testing was performed using \emph{historical} training data (without finetuning), and we selected the architecture that performed best on the validation dataset. 

We train the CNN using the Adam optimizer with a learning rate of $10^{-5}$ and a mean squared error loss function in PyTorch \citep{PyTorch}. We use early stopping with a patience of 20 epochs and select the model with the lowest validation loss. We train the CNN for a maximum of 1000 epochs, but it typically converges much earlier. To finetune the model on CFMIP data, we continue training on the CFMIP training dataset, with early stopping and the same learning rate and loss function as before.

%
%
\clearpage


\begin{figure}[ht]
    \centering
    \includegraphics{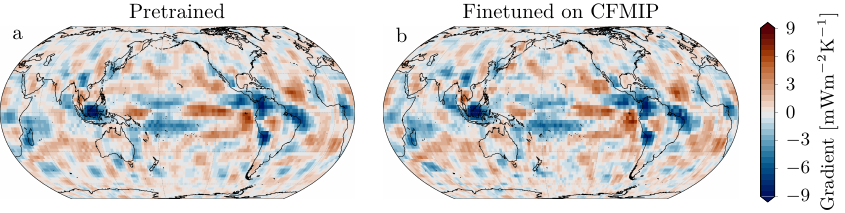}
    \caption{
        Gradient of the CNN, as the derivative of the output (global mean radiation $R$) to the input (near-surface temperature ${\bf T}$). The maps can be interpreted as a local radiative feedback. Panel a shows the gradient of the CNN pretrained on 7 large ensemble climate models. The gradient is averaged over all training years and models. Panel b shows the gradient of the CNN finetuned on CFMIP simulations, averaged over all CFMIP models and years in the training dataset. 
    }
    \label{fig:CNN_gradient}
\end{figure}

\clearpage
\begin{figure}[ht]
    \centering
    \includegraphics{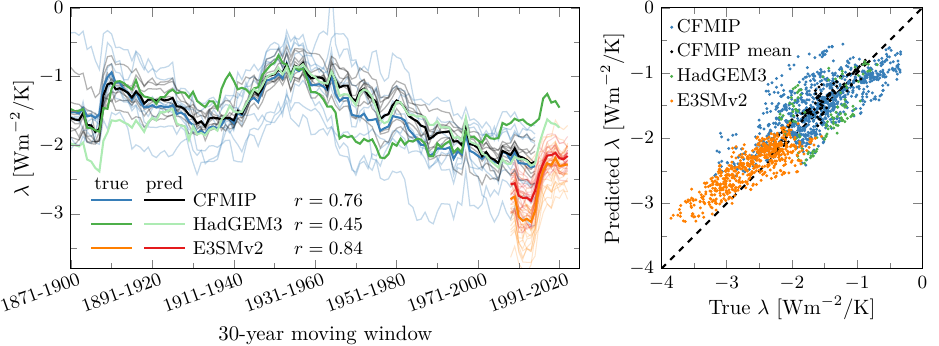}
    \caption{
        Performance of the CNN on AMIP model simulations. Lines labeled ``true'' are calculated directly from the model output. Lines labeled ``pred'' are calculated by applying the CNN to the models' \SI{2}{m} temperature output and then calculating $\lambda$. Values $r$ in the legend indicate the Pearson correlation between true and predicted time series for the given set of simulations across all models and ensemble members. The correlation between the multi-model CFMIP mean true and predicted time series is $r=0.96$. Right panel shows the true versus predicted $\lambda$ for all years in the AMIP simulations. The 1-to-1 line (dashed) indicates a perfect prediction. ``CFMIP3 mean'' (black dots) refers to the multi-model mean across CFMIP models (thick blue and black lines in left panel), indicating that the CNN tends to predict the ensemble mean $\lambda$ well. 
    }
    \label{fig:CNN_performance}
\end{figure}

\begin{figure}[ht]
    \centering
    \includegraphics{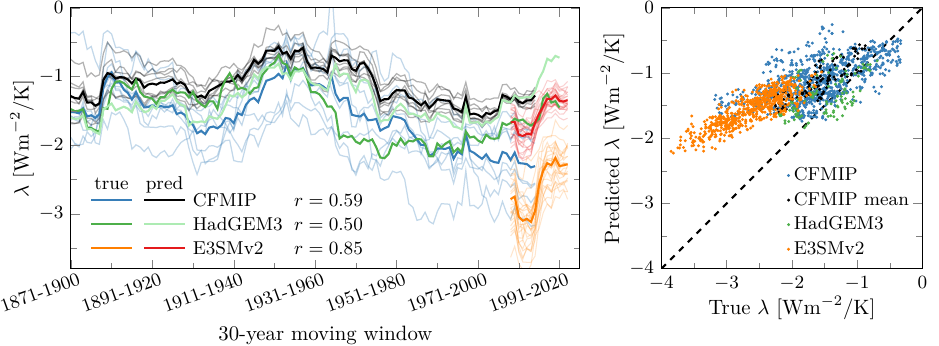}
    \caption{
        As Fig.~\ref{fig:CNN_performance}, but using the pretrained CNN without finetuning on CFMIP data. The correlation between the true and predicted $\lambda$ is still high (see legend; $r=0.86$ across the CFMIP multi-model mean), but the CNN tends to underestimate the magnitude of $\lambda$.
    }
    \label{fig:CNN_performance_pretrain}
\end{figure}

\begin{figure}[ht]
    \centering
    \includegraphics{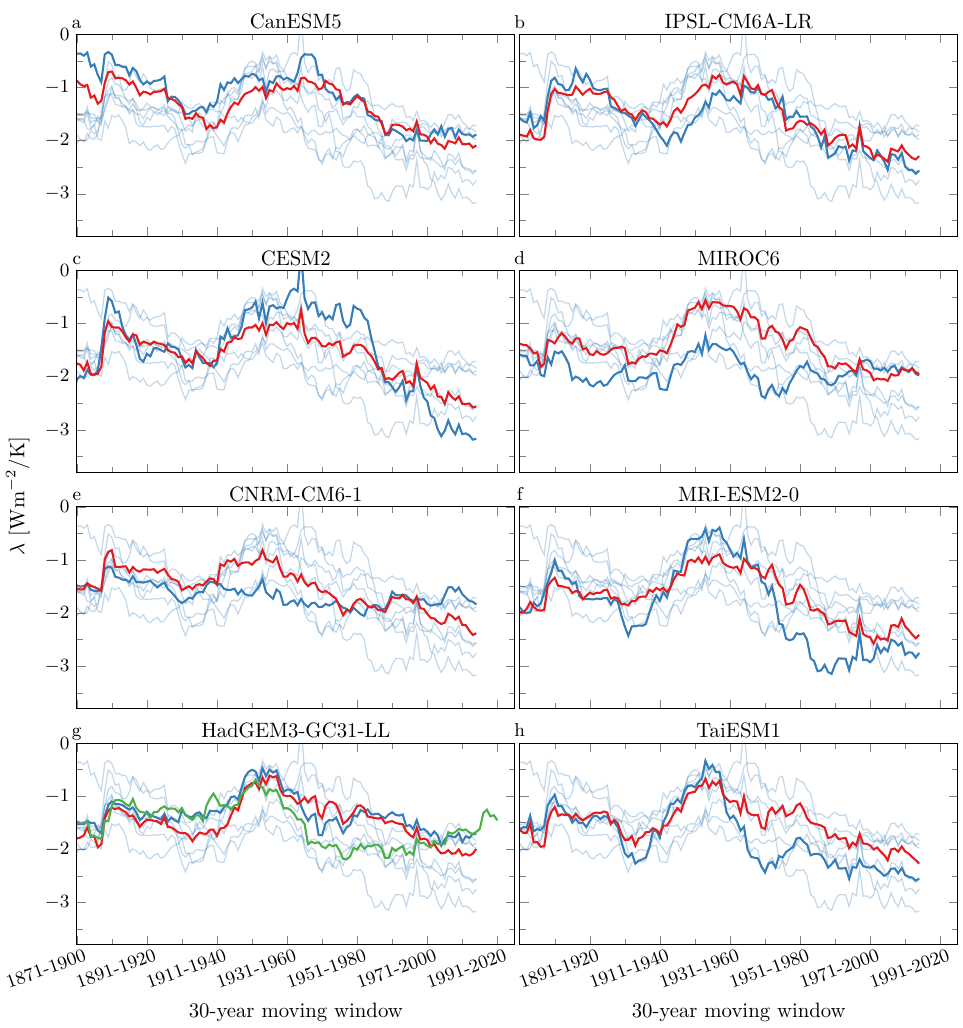}
    \caption{
        Model-dependence of the global radiative feedback parameter $\lambda$ in CFMIP simulations. Each panel shows results for a different model (thick blue), indicated at the top, as compared to the full CFMIP ensemble (thin blue). Red lines are calculated by applying the CNN to the models' \SI{2}{m} temperature output and then calculating $\lambda$. For HadGEM3-GC31-LL, we also include the new \emph{amip-piForcing} run in green.
    }
    \label{fig:historical_feedback_AMIP_models}
\end{figure}

\begin{figure}[ht]
    \centering
    \includegraphics[width=\textwidth]{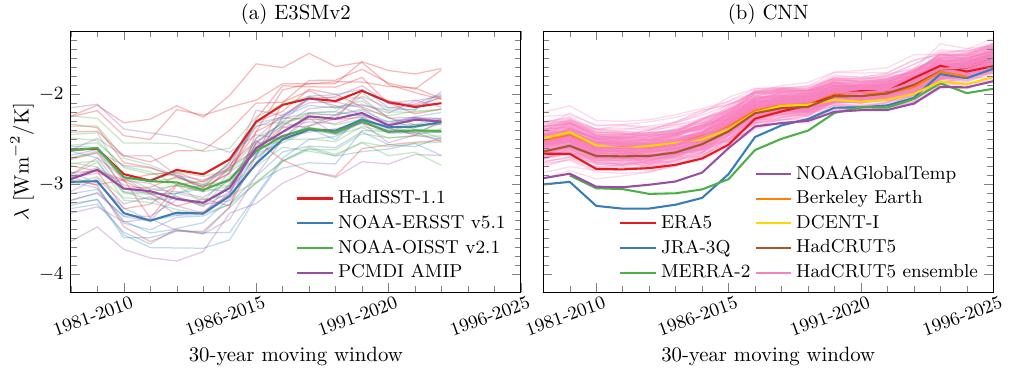}
    \caption{Dependence of the global radiative feedback $\lambda$ on surface temperature reconstructions. Panel a shows $\lambda$ calculated in the E3SMv2 ensemble, with ensemble means per SST boundary condition in thick lines. Thin lines show individual ensemble members. Panel b shows $\lambda$ calculated by applying the CNN to different temperature reconstructions. Pink thin lines show results from the CNN applied to the HadCRUT5 200-member ensemble, indicating observational and analysis uncertainty of the reconstruction \citep{Morice21}.}
    \label{fig:historical_feedback_datasets}
\end{figure}

\begin{figure}[ht]
    \centering
    \includegraphics[width=\textwidth]{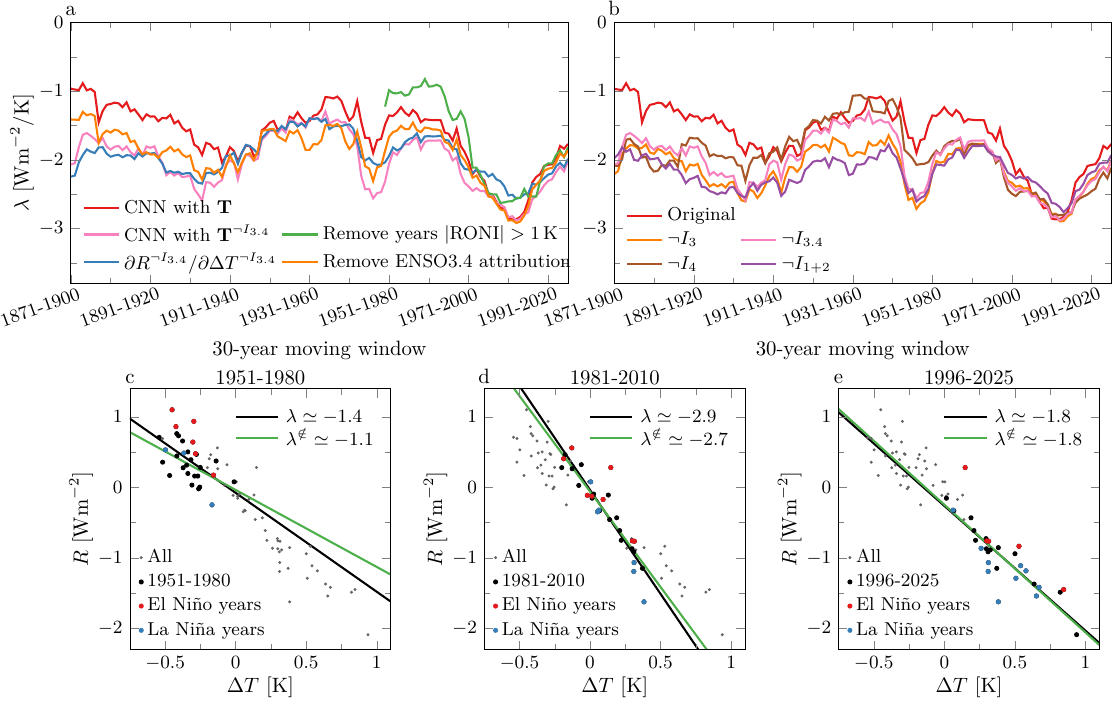}
    \caption{
        The effect of ENSO on the global radiative feedback $\lambda$. Panel a shows the feedback calculated by applying the CNN to ${\bf T}_{\rm rc}$ and different ways of removing the effect of ENSO. Only the average over all reconstructions is shown for visibility. 
        Red and pink lines are reproduced from Fig.~1b in the main text (original and feedback calculated by linearly regressing the Ni\~n{o} 3.4 index $I_{3.4}$ out of ${\bf T}_{\rm rc}$). 
        The blue line is calculated by regressing $I_{3.4}$ out of global mean temperature $\Delta T$ and CNN-predicted global mean radiation $R$ before calculating $\lambda$.
        The green line removes strong ENSO years from each 30-year period, defined as years with a October-November-December Relative Oceanic Ni\~{n}o Index \cite[RONI,][]{NOAA_CPC_RONI} smaller than $\SI{-1}{K}$ (strong La Ni\~{n}a) or larger than $\SI{1}{K}$ (strong El Ni\~{n}o). 
        The orange line is calculated by summing up the attribution of all regions but Ni\~{n}o 3.4 (i.e., as in Fig.~2k-p, but excluding the Ni\~{n}o 3.4 region). 
        Panel b shows the feedback calculated by the CNN applied to ${\bf T}_{\rm rc}$ with different ENSO indices regressed out (red and pink line are reproduced from the main text, other lines use Ni\~{n}o 3, Ni\~{n}o 4, and Ni\~{n}o 1+2 regions instead). 
        Panels c-e show the CNN-predicted $R$ as a function of $\Delta T$ in three different periods. Small gray dots show all values between 1950-2025, black dots the values in the 30-year period. Red dots show strong El Ni\~{n}o years (${\rm RONI}>\SI{1}{K}$), blue dots strong La Ni\~{n}a years (${\rm RONI}<\SI{-1}{K}$). Lines are calculated by regressing $R$ to $\Delta T$ for all years in the period (black) and for weak ENSO years only (excluding red and blue years; green). The slopes of the regression lines are the feedback $\lambda$ in that period (values given in \si{Wm^{-2}/K}). ENSO years do not stand out among other years in the 30-year window, and the regression lines are similar whether ENSO years are included or not, indicating that ENSO does not have a strong influence on the feedback $\lambda$ in these 30-year periods.
    }
    \label{fig:historical_feedback_enso_analysis}
\end{figure}

\begin{figure}[ht]
    \centering
    \includegraphics[width=\textwidth]{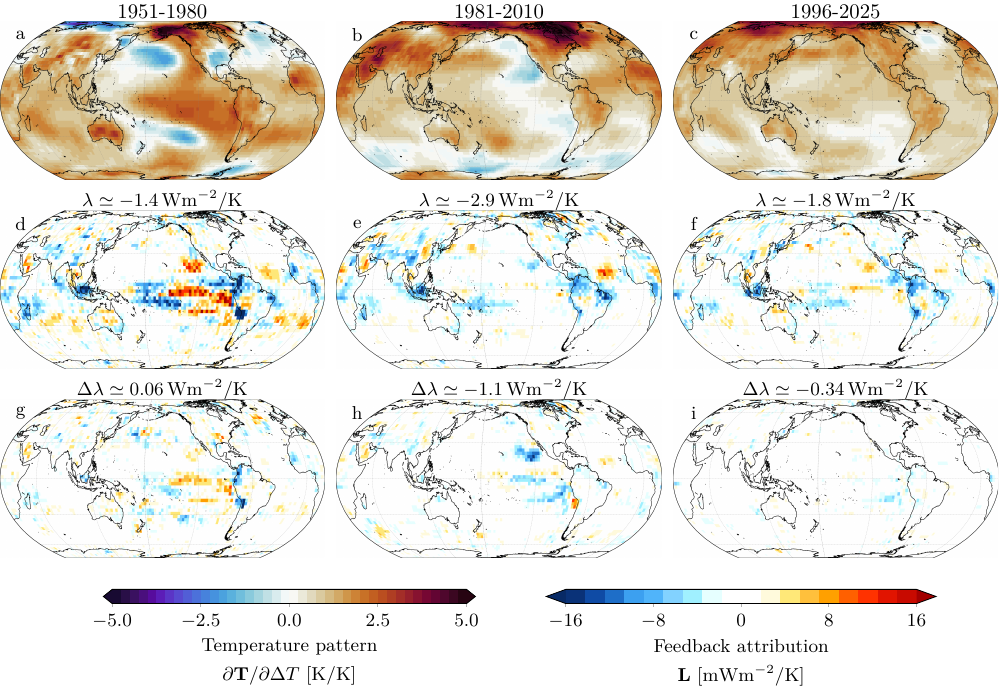}
    \caption{
        Attribution of changes in the global radiative feedback parameter, as in Fig.~2. Panels a-f are reproduced from Fig.~2a-f from the main text for reference. The attribution maps (panels d-f) give values that make $\lambda$ different from zero, because they are calculated with respect to a zero-everywhere uniform temperature pattern. That is, the sum of all grid boxes equates to $\lambda$. Panels g-i also show SHAP attribution maps, but now with respect to a uniform temperature pattern that has the same global mean as the actual temperature pattern. This allows us to isolate the effect of the spatial pattern of temperature change with respect to uniform temperature change. The sum of all grid boxes in panels g-i does not equate to $\lambda$, but rather to the difference  $\Delta\lambda = \lambda - \lambda_{\rm uniform}$ (indicated above the maps), with $\lambda_{\rm uniform}$ the feedback predicted by the CNN to a uniform temperature pattern.
    }
    \label{fig:shap_diffs_vs_uniform}
\end{figure}

\begin{figure}[ht]
    \centering
    \includegraphics{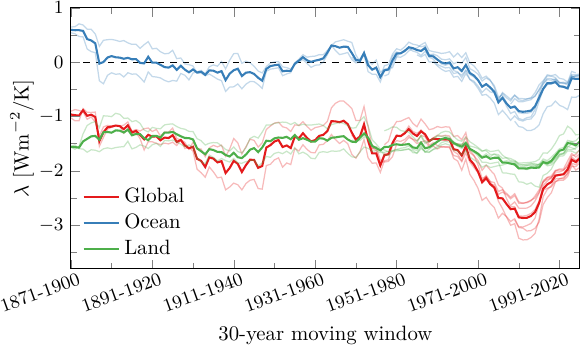}
    \caption{
        Attribution of changes in the global radiative feedback parameter to ocean and land surface temperature changes. As Fig.~3 in the main text, but showing absolute values instead of anomalies with respect to the long-time mean. The sum of ocean (blue) and land (green) contributions equals the global feedback (red, as in Fig.~1 of the main text). 
    }
    \label{fig:historical_feedback_attr_ocean}
\end{figure}

\begin{figure}[ht]
    \centering
    \includegraphics[width=\textwidth]{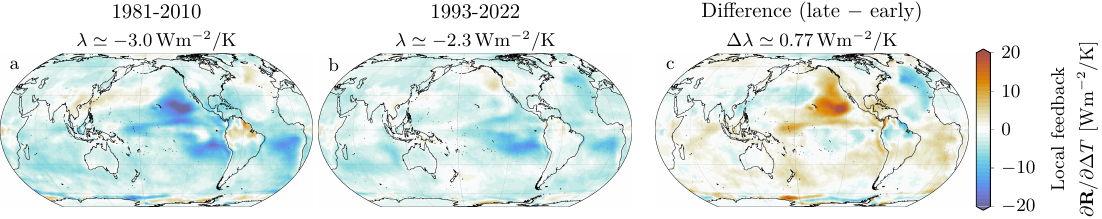}
    \caption{Local contribution to the global feedback, computed by regressing local $R_j$ in each grid box $j$ to global mean temperature $\Delta T$ in a given 30-year window in E3SMv2. These maps indicate how local $R_j$ varies with global mean temperature $\Delta T$, as opposed to attribution maps in the main text, which show how global mean $R$ varies with local $T_j$. The global mean of these maps equals the global radiative feedback $\lambda$.}
    \label{fig:dRidT_E3SM}
\end{figure}

\clearpage

\begin{table}[ht]
    \centering
    \caption{Overview of model data used in this study. Members marked with an asterisk are \emph{not} used in the training of the CNN.}
    \begin{tabular}{r | c c c l}
        Model & Simulation & \# & period & Reference \\\hline
        CanESM5 & \emph{historical} & 25 & 1871-2014 & \cite{Swart19} \\
            & \emph{piClim-histall} & 3 & 1871-2014 &  \\
            & \emph{amip-piForcing} & 1 & 1870-2014 &  \\\hline
        CNRM-CM6-1 & \emph{historical} & 26 & 1871-2014 & \cite{Voldoire19} \\
            & \emph{piClim-histall} & 1 & 1871-2014 &  \\
            & \emph{amip-piForcing} & 1 & 1870-2014 &  \\\hline
        E3SMv2 & \emph{historical} & 20 & 1871-2014 & \cite{Golaz22} \\
            & \emph{piClim-histall} & 3 & 1871-2014 &  \\
            & \emph{AMIP-F2010} & 10${}^\ast$ & 1979-2022 &  \\
            & \emph{NOAA-ERSST-v5-F2010} & 10${}^\ast$ & 1979-2022 &  \\
            & \emph{OISST-v2.1-F2010} & 10${}^\ast$ & 1979-2022 &  \\
            & \emph{HadISST-1.1-F2010} & 10${}^\ast$ & 1979-2022 &  \\
            & \emph{HadISST-1.1-F2010-NEP} & 5${}^\ast$ & 1979-2022 &  \\\hline
        HadGEM3-GC31-LL & \emph{historical} & 50 & 1871-2014 & \cite{Kuhlbrodt18} \\
            & \emph{piClim-histall} & 3 & 1871-2014 &  \\
            & \emph{amip-piForcing} & 1 & 1870-2014 &  \\
            & \emph{amip-piForcing} & 1${}^\ast$ & 1871-2020 &  \\\hline
        IPSL-CM6A-LR & \emph{historical} & 33 & 1871-2014 & \cite{IPSL-CM6A-LR} \\
            & \emph{piClim-histall} & 3 & 1871-2014 &  \\
            & \emph{amip-piForcing} & 1 & 1870-2014 &  \\\hline
        MIROC6 & \emph{historical} & 50 & 1871-2014 & \cite{MIROC6} \\
            & \emph{piClim-histall} & 3 & 1871-2014 &  \\
            & \emph{amip-piForcing} & 1 & 1870-2014 &  \\\hline
        MPI-ESM1.1 & \emph{historical} & 100 & 1871-2005 & \cite{MPI} \\
            & \emph{RCP8.5} & 100 & 2006-2014 & \\
            & \emph{piClim-histall} & 1 & 1871-2014 & \cite{Alessi23} \\
            & \emph{amip-piForcing} & 1 & 1870-2014 &  \\\hline
        CESM2 & \emph{amip-piForcing} & 1 & 1870-2014 & \cite{Danabasoglu20} \\\hline
        MRI-ESM2-0 & \emph{amip-piForcing} & 1 & 1870-2014 & \cite{Yukimoto19} \\\hline
        TaiESM1 & \emph{amip-piForcing} & 1 & 1870-2014 & \cite{Lee20} \\
    \end{tabular}
\end{table}

\begin{table}[ht]
    \centering
    \caption{Overview of observational and reanalysis data used in this study.}
    \begin{tabular}{r | c c l}
        Dataset & period & \#members & Reference \\\hline
        ERA5 & 1940-2025 & 1 & \cite{Hersbach20} \\
        JRA-3Q & 1948-2025 & 1 & \cite{Kosaka24} \\
        MERRA2 & 1980-2025 & 1 & \cite{Gelaro2017} \\
        DCENT-I & 1850-2025 & 1 & \cite{DCENTdata} \\
        NOAAGlobalTemp & 1850-2025 & 1 & \cite{NOAAGlobalTemp} \\
        Berkeley Earth${}^\ast$ & 1956-2024 & 1 & \cite{Rohde20} \\
        HadCRUT5.1.0.0${}^\dagger$ & 1979-2025 & 200 & \cite{Morice21} \\\hline
        DEEP-C & 1985-2000 & 1 & \cite{DEEPCdata} \\
        CERES-EBAF4.2.1 & 2001-2024 & 1 & \cite{CERESEBAF42} \\
        Effective radiative forcing & 1985-2024 & 1 & \cite{Forster25} \\
    \end{tabular}
    \begin{justify}
        ${}^\ast$We only use Berkeley Earth data starting in 1956, as earlier data has missing values.
        ${}^\dagger$Starting in 1979, HadCRUT5 has less than \SI{4}{\percent} of global grid cells missing. We fill these missing values using NOAAGlobalTemp data. Setting these values to climatology instead does not affect our results. Most of the missing values are located in the Southern Ocean, which contribute little to the CNN's predicted radiation.
    \end{justify}
\end{table}

\begin{table}[ht]
    \centering
    \caption{Hyperparameters used in testing different CNN architectures in Supporting Information Fig.~\ref{hpts}. All CNNs use a similar architecture as shown in Supporting Information Fig.~\ref{CNN_architecture}, but we change the  number of convolutional layers/kernels, kernel size, and amount of dense layers/nodes. The set of hyperparameters used in the main text is hpt0. 
    }
    \begin{tabular}[h]{r | c c c}
        & Convolutional layers & Dense layers & Learning rate\\\hline
        hpt0 & [32, 32]     & [32, 16] & $10^{-5}$\\
        hpt1 & [32, 32, 32] & [32, 16] & $10^{-5}$\\
        hpt2 & [32, 32]     & [32, 16] & $10^{-5}$\\
        hpt3 & [32, 32]     & [16, 8]  & $10^{-5}$\\
        hpt4 & [32, 32]     & [32, 16] & $10^{-4}$\\
        hpt5 & [16, 16]     & [32, 16] & $10^{-5}$\\
    \end{tabular}
    \label{tab:hpts}
\end{table}

%
%

\clearpage
\twocolumn
\setstretch{1.1}
\bibliographystyle{ametsocV6}

\end{document}